\documentclass[a4paper]{jpconf}
\usepackage{amsmath,amssymb,epsfig,bm,mathrsfs,color}
\usepackage{slashed}
\newcommand{\be}{\begin{equation}}
\newcommand{\ee}{\end{equation}}
\newcommand{\bea}{\begin{eqnarray}}
\newcommand{\eea}{\end{eqnarray}}

\newcommand{\vecp}{{\bm p}}
\newcommand{\vecP}{{\bm P}}
\newcommand{\veck}{{\bm k}}
\newcommand{\vecQ}{{\bm Q}}
\newcommand{\vecr}{\bm r}

\definecolor{red}{rgb}{0.8,0,0}
\definecolor{orange}{rgb}{0.8,0.2,0.0}
\definecolor{blue}{rgb}{0.3,0.0,0.8}

\begin{document}

\title{Inhomogeneous condensates in dilute nuclear matter and 
BCS-BEC crossovers}

\author{Martin Stein, Armen Sedrakian}
\address{Institute for Theoretical Physics, J.~W.~Goethe-University,
D-60438 Frankfurt-Main, Germany}
\ead{mstein@th.physik.uni-frankfurt.de, sedrakian@th.physik.uni-frankfurt.de}

\author{Xu-Guang Huang}
\address{Physics Department \& Center for Particle Physics and Field Theory, Fudan University, Shanghai 200433, China}
\ead{huangxuguang@fudan.edu.cn}

\author{John W. Clark}
\address{Department of Physics \& McDonnell Center for the Space Sciences, 
Washington University, St.~Louis, MO 63130, USA, and Center for 
Mathematical Sciences, University of Madeira, Funchal, Portugal}
\ead{jwc@wuphys.wustl.edu}

\author{Gerd R\"opke}
\address{Institut f\"ur Physik, Universit\"at Rostock,
  Universit\"atsplatz 3, D-18051 
Rostock, Germany}
\ead{gerd.roepke@uni-rostock.de}

\begin{abstract}
  We report on recent progress in understanding pairing phenomena in
  low-density nuclear matter at small and moderate isospin
  asymmetry. A rich phase diagram has been found comprising various
  superfluid phases that include a homogeneous and phase-separated BEC
  phase of deuterons at low density and a homogeneous BCS phase, an
  inhomogeneous LOFF phase, and a phase-separated BCS phase at higher
  densities. The transition from the BEC phases to the BCS phases is
  characterized in terms of the evolution, from strong to weak
  coupling, of the condensate wavefunction and the second moment of
  its density distribution in $r$-space. We briefly discuss approaches
  to higher-order clustering in low-density nuclear matter.
\end{abstract}

\section{Introduction}\label{sec:1}

The two-nucleon interactions in nuclear matter at sub-saturation
densities (below about half the saturation density) are well
constrained by the phase-shift data and analysis of elastic
nucleon-nucleon collisions, facilitating a quantitative many-body
description of nuclear matter in this regime.  However, a significant
theoretical challenge is posed by the complexity of behavior arising
from the dominant attractive part of the nuclear interaction, notably
the formation of nuclear clusters and the emergence of nucleonic pair
condensates of the Bardeen-Cooper-Schrieffer (BCS) type.  Such a
description is of direct relevance for the matter found in supernovae
and neutron stars.  These two settings differ somewhat in the regions
of parameter space (density, temperature, isospin asymmetry) involved.
For example, in supernovae the nuclear matter is at finite isospin
asymmetry which, however, is small compared to that of cold 
$\beta$-catalyzed neutron star matter.  In this contribution we review some
recent progress in establishing the nature of pairing in low-density
nuclear matter.

\section{Pairing}\label{sec:2}
If we restrict ourselves to only two-body correlations,
nuclear matter at extremely low density is a mixture of
quasi-free nucleons and deuterons, 
the only effect of the interaction being
to renormalize the mass of the constituents. 
As bosons, the deuterons
may also condense and form a Bose-Einstein
condensate (BEC) even without interactions, provided, of course, the
temperature is sufficiently
low~\cite{Alm:1993zz,Baldo:1995zz,Sedrakian:2005db,Mao:2008wz,Huang:2010fk,Jin:2010nj}.
Increasing the density of the system while keeping its temperature 
constant serves to increase the degeneracy of the system.  As the density
$\rho$ approaches the saturation density $\rho_0 =
2.8\times 10^{14}$g cm$^{-3}$ of symmetrical nuclear matter, 
the abundance of deuterons is reduced by
Pauli blocking of the phase space available to nucleons~(for a
recent discussion see~\cite{Ropke:2012qv}).  
However, because at high densities the nucleons fill a Fermi sphere and 
because the interaction between nucleons is attractive, there emerges
a BCS-type coherent state -- a condensate of paired nucleons.

The dominant pairing interaction comes from the $^3S_1$-$^3D_1$ ($SD$)
partial wave, i.e., the same interaction channel that binds the
deuteron. Therefore, we may anticipate that the nuclear matter
undergoes a BEC-BCS phase transition as the density increases. This
effect has been conjectured to occur in the context of
intermediate-energy heavy-ion
collisions~\cite{Alm:1993zz,Baldo:1995zz} and more recently in the
context of supernovae~\cite{Heckel:2009br,Stein:2012wd}.

The theoretical framework developed by  Nozi\`eres and 
Schmitt-Rink~\cite{Nozieres:1985zz} for description of the BCS-BEC transition in
the condensed-matter context can be transposed to nuclear matter.
Importantly, however, isoscalar neutron-proton ($np$) pairing is
disrupted by the isospin asymmetry induced by weak interactions in
stellar environments and expected in exotic nuclei, since the mismatch
in the Fermi surfaces of protons and neutrons suppresses the pairing
correlations~\cite{Sedrakian:1999cu}.  The standard
Nozi\`eres-Schmitt-Rink theory of the BCS-BEC crossover has been
modified to account for this fact~\cite{Lombardo:2001ek}.
\begin{figure}[tb]
\begin{center}
\includegraphics[width=11.5cm,height=9cm]{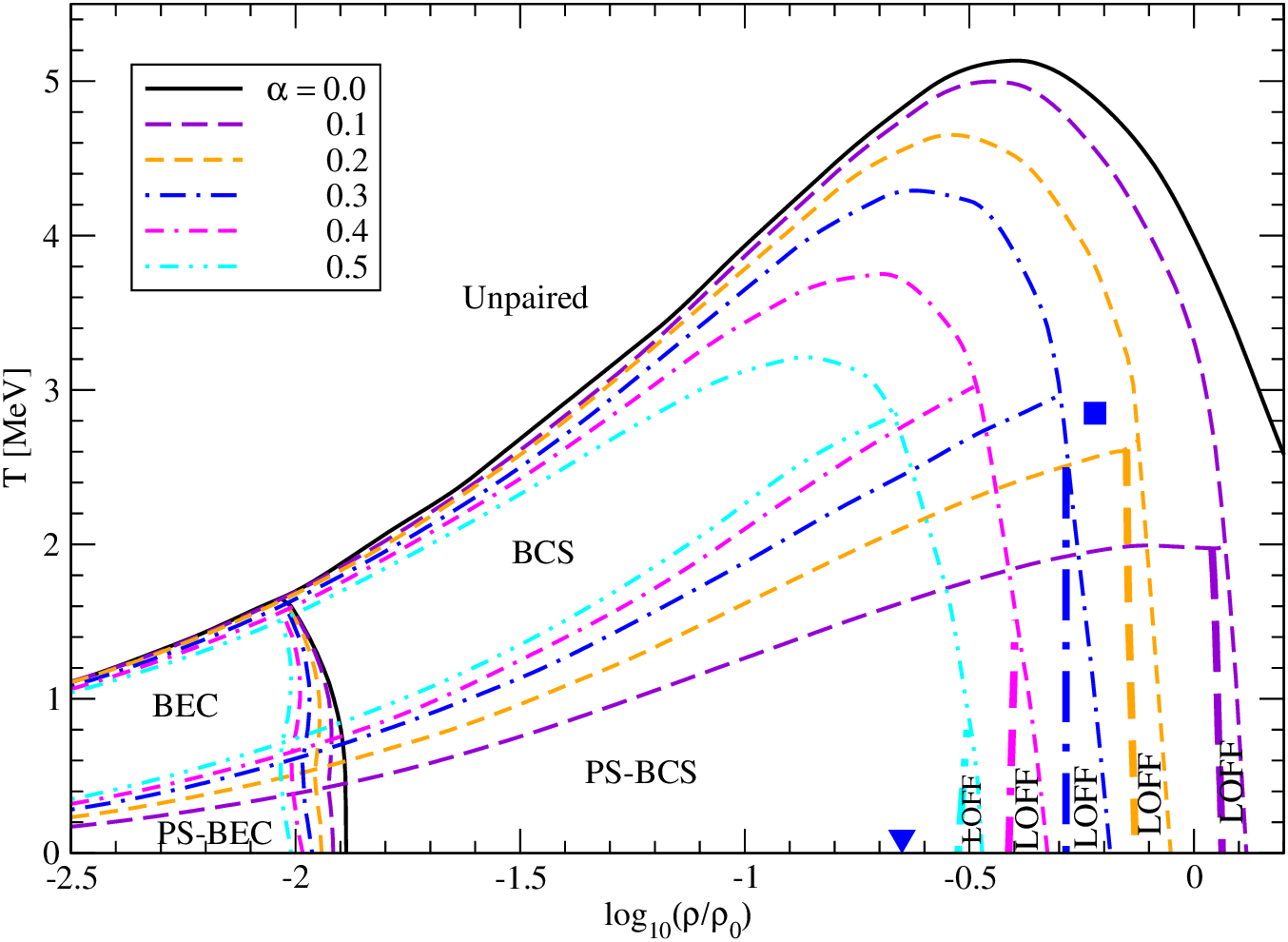}
\caption{Phase diagram of dilute nuclear matter in the
  temperature-density plane for several isospin asymmetries $\alpha$
  (from~\cite{Stein:2012wd}).  Included are four phases: unpaired
  phase, BCS (BEC) phase, LOFF phase, and PS-BCS (PS-BEC) phase. For
  each asymmetry there are two tri-critical points, one of which is
  always a Lifshitz point~\cite{Chaikin}. For special values of
  asymmetry these two points degenerate into single tetra-critical
  point at $\log(\rho/\rho_0) = -0.22$ and $T = 2.85$ MeV (shown by a
  square dot) for $\alpha_4 = 0.255$. The LOFF phase disappears at the
  point $\log(\rho/\rho_0) = -0.65$ and $T=0$ (shown by a triangle)
  for $\alpha = 0.62.$ The boundaries between BCS and
  BEC phases are identified by the change of 
 sign of the average chemical potential $\bar\mu$.
}
\label{fig:2}
\end{center}
\end{figure}
The emergence of two mismatched Fermi surfaces in isospin-asymmetrical
nuclear matter introduces a new scale into the problem, namely the
shift $\delta\mu = (\mu_n - \mu_p)/2$ in the chemical potentials
$\mu_n$ and $\mu_p$ of neutrons and protons (up or down) from their
otherwise common value $\bar\mu = (\mu_n + \mu_p)/2$.  If sufficiently
large, this shift can destroy the condensate.

In the isospin symmetrical case (at $\delta\mu=0$) the condensate is
characterized by a gap $\Delta_0$ in the $SD$ channel of order several
MeV.  A sequence of unconventional phases has been conjectured to
occur with increasing isospin symmetry, i.e., as $\delta\mu$ increases
from zero to values of order $\Delta_0$.  One of these is a
neutron-proton condensate whose Cooper pairs have non-zero
center-of-mass (CM) momentum~\cite{Sedrakian:2000an,Muther:2002dm,Shang:2013cma}.
This phase is the analogue of the Larkin-Ovchinnikov-Fulde-Ferrell
(LOFF) phase in electronic superconductors~\cite{LO,FF}. Another
possibility is phase separation into superconducting and normal
components, proposed in the context of cold atomic
gases~\cite{Bedaque:2003hi}.  At large isospin asymmetry, where $SD$
pairing is strongly suppressed, a BCS-BEC crossover may also occur in
the isotriplet $^1S_0$ pairing channel. Finally, it has been
conjectured that spontaneous deformation of Fermi surfaces is an
alternative to the LOFF phase and can in fact dominate the latter in
nuclear systems~\cite{Muther:2002dm}.

Recently, the concepts of unconventional $SD$ pairing and the BCS-BEC
crossover were unified in a model of isospin-asymmetric nuclear
matter~\cite{Stein:2012wd} by including some of the phases mentioned
above.  A phase diagram for superfluid nuclear matter was constructed
over wide ranges of density, temperature, and isospin asymmetry. The
coupled equations for the gap and the densities of constituents
(neutrons and protons) were solved allowing for the ordinary BCS
state, its low-density asymptotic counterpart BEC state, and two
phases that owe their existence to the isospin asymmetry: the phase
with a moving condensate (LOFF phase) and the phase in which the
normal fluid and superfluid occupy separate spatial domains.  The
latter phase is referred to as the phase-separated BCS (PS-BCS) phase
and, in the strong-coupling regime, the phase-separated BEC (PS-BEC)
phase. In this phase the asymmetry is accumulated in the normal
domains, whereas the superfluid domain is perfectly isospin symmetric.

The gap equation to be solved has the form \bea \label{eq:gap}
\Delta_l(k,Q) &=&\frac{1}{2} \sum_{l'}\int\!\!\frac{d^3k'}{(2\pi)^3}
V_{l,l'}(k,k') \sum_{a,r}
\frac{\Delta_{l'}(k',Q)}{2\sqrt{E_{S}(k')^2+\Delta_{l'}^2(k',Q)}}[1-2f(E^r_a)],
\eea where the gap $\Delta(Q)$ in the quasiparticle spectrum is a
function of the modulus $Q$ of the total momentum of the pair of
particles, the $l$ summation is over the coupled partial waves,
$V(\veck,\veck')$ is the neutron-proton interaction potential,
$f(x)=1/[\exp{(x/T)}+1]$, $ E_{r}^{a} = [E_S^2+\Delta^2]^{1/2} +
r\delta\mu +a E_A, $ and $a, r \in \{+,-\}$, with $E_S
=\left(Q^2/4+k^2\right)/2m^*-\bar\mu$, $E_A = \veck\cdot \vecQ /2m^*$,
and $\Delta^2=\sum_l\Delta_l^2$.  Note that the gap in the $SD$ state
is  angle averaged in the denominator of the gap equation; this
approximation can be avoided~\cite{Shang:2013cma,Shang:2013wza}.
Here $\bar\mu$ is the average of the
chemical potentials $\mu_n$ and $\mu_p$ of neutrons and protons
introduced above, and $m^*$ is the effective mass of the
nucleons. (Note that we do not distinguish between the effective
masses of neutrons and protons.)

The resulting phase diagram of dilute nuclear matter is shown in
Fig~\ref{fig:2} for several values of isospin asymmetry $\alpha =
(n_n-n_p)/(n_n+n_p)$, where $n_{n}$ and $n_p$ are the number densities
of neutrons and protons.  Four different phases of matter are present.
(a) The unpaired normal phase is always the ground state at
sufficiently high temperatures $T>T_{c0}$, where $T_{c0}(\rho)$ is the
critical temperature of the normal/superfluid phase transition at
$\alpha =0$. (b) The LOFF phase is the ground state in a narrow
temperature-density strip at low temperatures and high densities. (c)
The domains of phase separation (PS) appear at low temperatures and
low densities, while the isospin-asymmetric BCS phase is the ground
state at intermediate temperatures for densities above the crossover
to a BEC phase and below the density where the condensate vanishes in
the isospin symmetrical limit.

In the extreme low-density and strong-coupling regime, the BCS 
superfluid phases have two counterparts: the BCS phase evolves into 
the BEC phase of deuterons, whereas the PS-BCS phase evolves into the 
PS-BEC phase, in which the superfluid fraction of matter is a BEC of 
deuterons.  The superfluid/unpaired phase transitions and the phase 
transitions between the superfluid phases are of second order (thin 
solid lines in Fig.~\ref{fig:2}), with the exception of the PS-BCS 
to LOFF transition, which is of first order (thick solid lines in
Fig.~\ref{fig:2}). The BCS-BEC transition and the PS-BCS to PS-BEC
transition are smooth crossovers.  At non-zero isospin asymmetry the
phase diagram features two tri-critical points where the simpler
pairwise phase coexistence terminates and three different phases
coexist.

Consistent with the earlier studies of the BCS-BEC crossover, one
observes in the phase diagram of Fig.~\ref{fig:2} a smooth crossover
to an asymptotic state corresponding to a mixture of a Bose condensate
of deuterons and a gas of excess neutrons. This however occurs at
moderate temperatures, where the unconventional phases do not appear.
The new ingredient of the nuclear phase diagram is the crossover seen
at very low temperatures, where the heterogeneous superfluid phase is
replaced by a heterogeneous mixture of a phase containing a deuteron
condensate and a phase containing neutron-rich unpaired nuclear
matter.

The transition to the BEC regime of strongly-coupled neutron-proton pairs, 
which are asymptotically identical with deuterons, occurs at low densities.  
The criterion for the transition from BCS to BEC is that either the average 
chemical potential $\bar \mu$ changes its sign from positive to negative 
values, or the coherence length $\xi$ of a Cooper pair becomes 
comparable to the interparticle distance, 
i.e., $\xi$ becomes of order $d\sim \rho^{-1/3}$ as it ranges
from $\xi \gg d$ to $\xi \ll d$.  The coherence length is related to 
the wave function of the condensate $\Psi(\vecr)$ 
by 
\be \label{eq:xi}
\xi = \sqrt{\langle r^2\rangle}, \quad \quad \langle r^2\rangle= 
\int d^3r 
\,
r^2 \vert
\Psi(r)\vert^2, 
\ee
where the wave-function is defined in terms of the kernel of the gap equation
according to 
\be
\Psi(\vecr) = \sqrt{N} \int \frac{d^3p}{(2\pi)^3}
[K(\vecp,\Delta)-K(\vecp,0)]e^{i\vecp\cdot\vecr},
\qquad \int d^3r \vert \Psi(\vecr)\vert^2 = N^{-1}.
\ee
Thus the change in the coherence length is
related to the change of the condensate wave-function across the
BCS-BEC crossover. In the case of neutron-proton pairing, the criteria
for the BCS-BEC transition are fulfilled, i.e., $\bar \mu$ changes
sign and the mean distance between the pairs becomes larger than the
coherence length of the superfluid.
\begin{figure}[tb]
\begin{center}
\includegraphics[width=11.5cm,height=8cm]{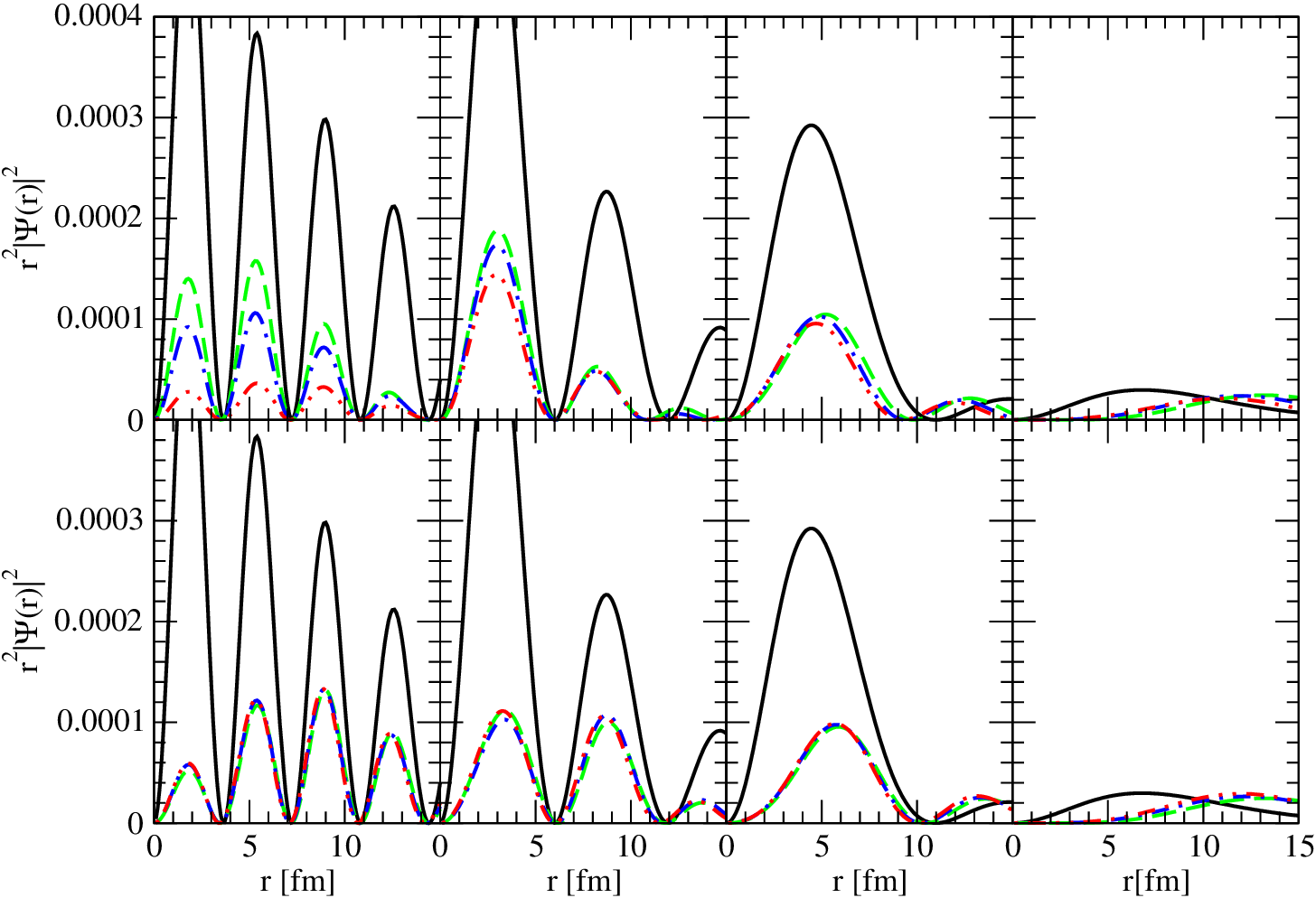}
\caption{ The quantity $r^2|\Psi(r)|^2$ as a function of radial
  coordinate $r$ for $T = 0.5$ MeV and densities (from left to right)
  $\log_{10}\rho/\rho_0 = -0.5, -1., -1.5, -2.5$ plotted for the
  asymmetrical BCS state (upper panel) and the PS phase (lower
  panel). The values of asymmetry are $\alpha = 0.0$ (solid lines),
  0.1 (dashed lines), 0.2 (dash-dotted lines), and 0.3
  (dash-double-dotted lines).
\vspace{1.cm}
 }
\label{Psi}
\end{center}
\end{figure}
\begin{figure}[tb]
\begin{center}
\includegraphics[width=11cm,height=8cm]{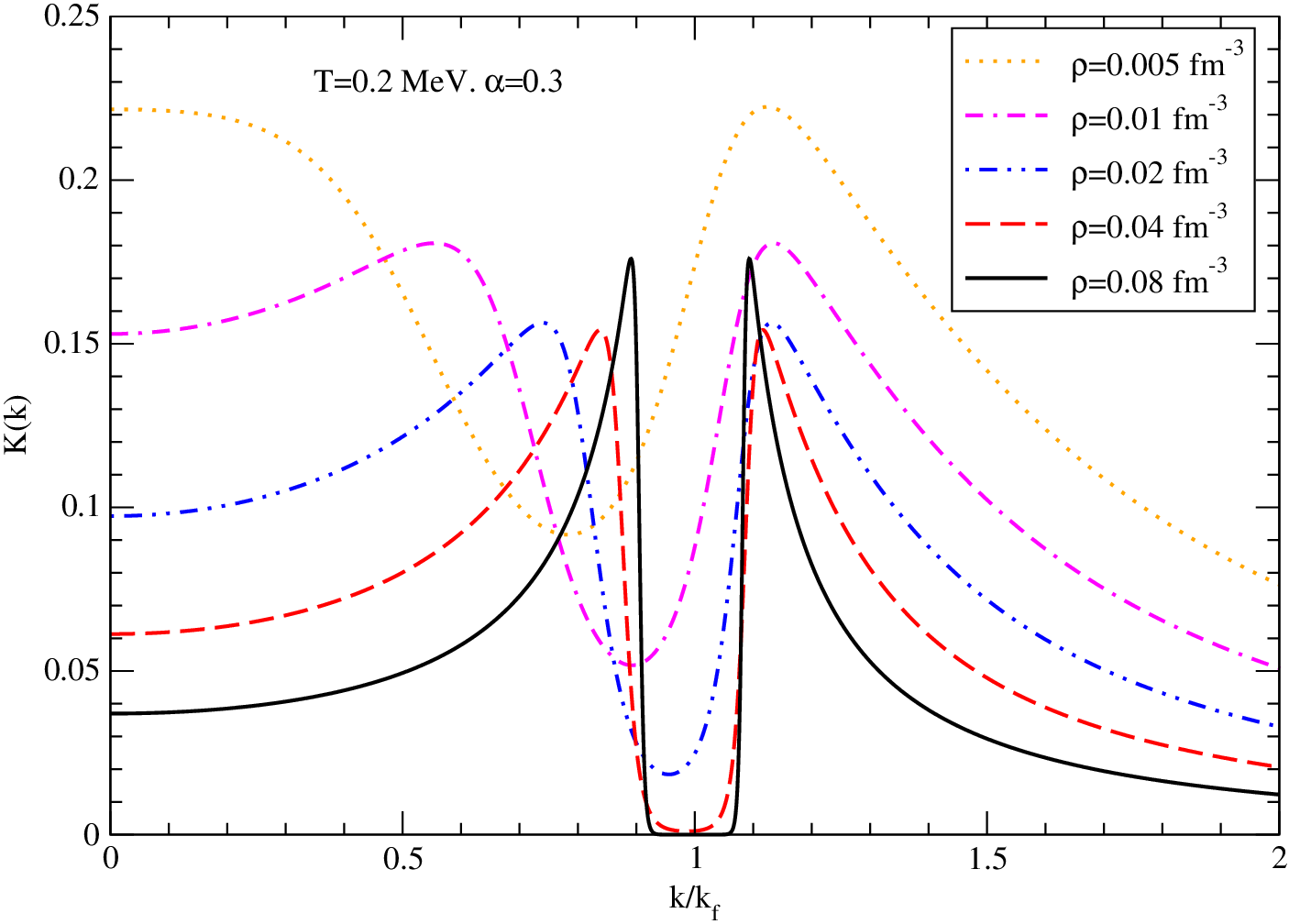}
\caption{ Kernel $K(k)$ of the gap equation (\ref{eq:gap}) as a
 function of momentum, in units of the relevant Fermi momentum $k_F$
 for fixed $\alpha = 0.3$ and $T=0.2$ MeV at various densities.}
\label{Psi2}
\end{center}
\end{figure}

In Fig.~\ref{Psi} we display the integrand $r^2|\Psi(r)|^2$ of
Eq.~(\ref{eq:xi}) as a function of radial distance for BCS and PS
states at various densities.  In its construction, the PS state
corresponds to the symmetrical BCS state insofar as we are concerned
with the nature of the Cooper pairs; hence there are no effects from
the asymmetry in this case; the asymmetry affects the PS condensate
indirectly via the number of particles available for pairing. In the
weak-coupling (high-density-limit) case, the integrand function has an
oscillatory form which extends over many periods of the interparticle
distance.  This behavior reflects the coherence associated with the
condensate, in which the spatial correlations are characterized by
scales much larger than the interparticle distance. For smaller
densities (larger couplings) the integrand of $\langle r^2 \rangle$ is
increasingly concentrated at the origin, with at most a few periods of
oscillation.  At low density (strong coupling) the pairs are well
localized in space within a radius which is small compared to the
interparticle distance; there is only one spike in the integrand
function.  This regime has BEC character, in that the pair
correlations just extend over distances comparable to the
interparticle distance. In the case of the BCS superfluid it is seen
that isospin asymmetry affects only the amplitude of the oscillations,
the period being unaffected in the weak-coupling limit and defined by
the magnitude of the inverse Fermi momentum.

Further insight into the BCS-BEC crossover can be gained from
examining the kernel $K(p)$ of the gap equation (\ref{eq:gap}), which
is defined by the sum appearing to the right of the interaction $V$ in
this equation.  Physically, $K(p)$ can be interpreted as the wave
function of the Cooper pairs, since it obeys a Schr\"odinger-type
eigenvalue equation in the limit of extremely strong coupling.  The
prefactor of the Pauli operator $1-2f$ is a smooth function of
momentum with a maximum at the Fermi surface, where $E_S$
vanishes. The momentum ranges whose contributions are important in the
gap equation in different regimes of the phase diagram can be
identified from Fig.~\ref{Psi2}.  In the BCS regime, $K(p)$ has two
sharp maxima which are separated by a depression (of width $\sim
\delta\mu$) around the Fermi momentum. Because of strong localization
in momentum space, the Cooper pairs have an intrinsic structure which is broad in
real space, implying a large coherence length.  This is characteristic
of the BCS regime. The picture is reversed in the strong-coupling
(low-density) limit, where $K(p)$ is a broad function of momentum,
corresponding to bound states (deuterons) well-localized in the real
space.  This is characteristic of the BEC regime.  In addition, as the
density decreases, the lower peak moves toward $k=0$, due to the fact
that $\bar{\mu}$ changes its sign from positive to negative at the
transition from the BCS to the BEC regime. As a consequence, the
prefactor of the Pauli operator $1-2f$ peaks at $k=0$ in the BEC
regime, rather at the Fermi surface as in the BCS regime.

\section{Higher-order clustering}

In supernovae the matter is at finite isospin asymmetry (but with
small values compared to those of cold neutron stars) and at
relatively high temperatures.  These conditions could allow for a
substantial presence of light clusters, notably deuterons, tritons,
$^3$He nuclei, and alpha
particles~\cite{Heckel:2009br,Sumiyoshi:2008qv,Typel:2009sy,Hempel:2009mc,Raduta:2010ym,Oertel:2012qd,Gulminelli:2011hr}. The
composition of nuclear matter in thermodynamic equilibrium at low
density is given by the mass action law, the so-called nuclear
statistical equilibrium (NSE).  At given temperature $T$ and total
baryonic density $\rho$, the partial densities of the components $c$
(whether nucleons or nuclei) are determined by a phase-space sum
(integral) over their Fermi or Bose distribution functions as
\begin{equation}
\label{compos}
n_c=g_c\sum_P \Bigg\{\exp\left[{\left(\frac{P^2}{2A_c m} +E_c-Z_c
  \mu_p-N_c\mu_n\right)}\Big/{T}\right]\pm 1
\Bigg\}^{-1}, 
\end{equation}
where $\vecP$ is the center-of-mass momentum, $Z_c$ and $N_c$ are the
proton and neutron numbers of component $c$, $A_c = N_c+Z_c$ is its
mass number, and $g_c$ is its degeneracy factor (accounting for spin
and excitations).  A number of
studies~\cite{Heckel:2009br,Sumiyoshi:2008qv,
  Typel:2009sy,Hempel:2009mc,Raduta:2010ym,Oertel:2012qd,Gulminelli:2011hr}
clearly demonstrate the importance of two- and four-body correlations.
The NSE approach is valid at low-densities, where the free-space
binding energies of bound states are not affected by the nuclear
environment.  As the baryon density increases, the energies of bound
states are modified due to the strong and electromagnetic
interactions. These modifications can be incorporated by solving an
``effective'' Schr\"odinger equations for an $A$-particle cluster (see
~\cite{Ropke2011} and references therein)
\begin{equation}
\label{inmedSGl}
\left[ \sum_{i=1}^A E_i^{\rm mf} -E\right]\psi(1\dots A)+
\sum_{i < j}^A[1-f(i)-f(j)]\sum_{1'\dots A'}V_{ij}(1\dots A,1' \dots
A')\psi(1' \dots A')=0\,,
\end{equation}
where the indices $i,j$ label nucleons,
$\psi(1 \dots A)$ is a bound-state wave function, $E$ is
its corresponding energy eigenvalue, $V_{ij}$ is the pairwise 
interaction potential, and $f(i)$ is the distribution function of 
nucleon $i$. 
The mean-field corrections to the quasiparticle energies are included
in $E^{\rm mf}_1$ and can be evaluated from the Skyrme or the
relativistic mean-field functionals. The modifications of energies of
higher-order bound states depend on the center-of-mass momentum
$\vecP$ as well as the thermodynamic parameters of the medium. These
have been computed in detail recently in Ref.~\cite{Ropke2011}, where
fitting formulas are also given. The main physical effect of these
modifications is to suppress the abundances of clusters
upon asymptotic approach to the nuclear saturation density.

To demonstrate the utility of the ``effective'' Schr\"odinger equation
(\ref{inmedSGl}), we consider the example of a four-nucleon
($\alpha$-like) cluster and assume for simplicity that both $T$ and
$\alpha$ are zero.  Bound states can appear below the continuum of
scattering states, whose edge is given simply by the sum of the
energies $E_F$ of two neutrons and two protons at the tops of their
respective Fermi seas.  In the zero-density limit, $\alpha$ particles
are formed with bound-state energy $E^0_\alpha=-28.3$ MeV.  The energy
of the center-of-mass motion vanishes at ${\vecP}=0$. At finite
density of the surrounding nuclear matter, the energy of the
four-nucleon bound state is shifted due to Pauli blocking, so that
$E_4[T,\rho]=E^0_\alpha+\Delta E^{\text{Pauli}}_4(T,\rho)\,.$ The
shift $E^{\text{Pauli}}_4$ has been found from Eq.~(\ref{inmedSGl})
within a variational approach using a Gaussian form for the wave
function and a Gaussian separable interaction, adjusted to reproduce
the binding energy and the r.m.s.\ radius of the $\alpha$
particle~\cite{Roepke1}.  The density dependence of the shift (at
vanishing $T$ and $\alpha$) is well approximated by $\Delta E_4^{\rm
  Pauli}(\rho) =4515.9\, \rho -100935\, \rho^2+1202538\, \rho^3$,
where the energy is given in units of MeV and the density in units of
fm$^{-3}$.  Using this formula, it is easy to deduce that the binding
energy of an $\alpha$ particle crosses the continuum $4 E_F$ at a
density $\rho \sim 0.03$ fm$^{-3}$, above which no $\alpha$ particle
bound state can be formed.  Analogous fits to in-medium binding
energies for other light clusters can be found
in~\cite{Typel:2009sy}. These compare well with exact solutions of the
three- and four-body quantum mechanical problems in the medium (see,
for example,
~\cite{Sedrakian:2005db,Sedrakian:1997zd,Beyer:2000ds} and
references therein).  A useful feature of Eq.~(\ref{compos})
is that it has the character of a Bose distribution function of a
cluster for even $A$ which implies that the physics of BEC of clusters
(notably the critical temperature) can be deduced starting from NSE or
its modifications that include density- and temperature-dependent
binding energies.

\section{Outlook}

It is apparent from the discussion above that 
low-density and moderately warm nuclear matter can have a rather 
complicated phase diagram at finite, but small, isospin asymmetries. 
It is also apparent that higher-order clusters need to be included 
in the phase diagram to make the discussion complete. Furthermore,
the $\alpha$ particles will form a Bose-Einstein condensate
at sufficiently low temperatures (see
\cite{Clark:1966,Sedrakian:2004fh,Yamada:2011bi} and references therein).
The extent to which these clusters modify the structure of the phase
diagram developed here remains to be studied.  The diverse microscopic
aspects of superfluid, asymmetrical nuclear matter portend diverse
ramifications for the astrophysics of supernovae and hot compact
stars.

\ack The work of MS was supported by the HGS-HIRe graduate program at
Frankfurt University.  XGH acknowledges support through Fudan
University grants EZH1512519 and EZH1512600.  JWC acknowledges
research support from the McDonnell Center for the Space Sciences and
expresses his appreciation for the hospitality of the Center of
Mathematical Sciences, University of Madeira.

\end{document}